\documentclass[submission,copyright,creativecommons]{eptcs}
\providecommand{\event}{SOS 2023} 

\usepackage{iftex}

\ifpdf
  \usepackage{underscore}         
  \usepackage[T1]{fontenc}        
\else
  \usepackage{breakurl}           
\fi
\usepackage{graphicx}

\title{The Way We Were: 
Structural Operational Semantics Research in Perspective}
\author{Luca Aceto 
\institute{Department of Computer Science, \\
Reykjavik University, \\
Reykjavik, Iceland}
\institute{Gran Sasso Science Institute, \\
L'Aquila, Italy}
\email{luca@ru.is \qquad luca.aceto@gssi.it}
\and 
Pierluigi Crescenzi
\institute{Gran Sasso Science Institute, \\
L'Aquila, Italy}
\email{pierluigi.crescenzi@gssi.it}
\and 
Anna Ing\'olfsd\'ottir
\institute{Department of Computer Science, \\
Reykjavik University, \\
Reykjavik, Iceland}\email{annai@ru.is}
\and
Mohammad Reza Mousavi
\institute{Department of Informatics, \\
King's College London\\
London, UK}
\email{mohammad.mousavi@kcl.ac.uk}
}
\def\titlerunning{SOS Research in Perspective}
\def\authorrunning{Aceto, Crescenzi, Ing\'olfsd\'ottir, and Mousavi}
\begin{document}
\maketitle

\begin{abstract}
\end{abstract}

\section{Introduction}\label{Sect:intro}

This position paper on the (meta-)theory of Structural Operational Semantic (SOS) is motivated by the following two questions:
\begin{itemize}
    \item Is the (meta-)theory of SOS dying out as a research field?
    \item If so, is it possible to rejuvenate this field  with a redefined purpose? 
\end{itemize}
In this article, we will consider possible answers to those questions by first analysing the history of the EXPRESS/SOS workshops and the data concerning the authors and the presentations featured in the editions of those workshops as well as their subject matters.

The first International Workshop on Structural Operation was held in London, UK in 2004. 
The  workshop was established as `a forum
for researchers, students and practitioners interested in new developments,
and directions for future investigation, in the field of structural operational
semantics. One of the specific goals of the workshop was to establish synergies
between the concurrency and programming language communities working on
the theory and practice of SOS.' At its ninth edition, the SOS workshop joined forces with the nineteenth edition of International Workshop on  Expressiveness in Concurrency. The joint workshop was meant to cover the broader scope of `the formal semantics of systems and programming concepts, and on the expressiveness of mathematical models of computation.'

We examined the contributions dedicated to the theory of SOS presented in the EXPRESSS/SOS workshop series (and, prior to that, in the SOS workshop) and whether they appeared before or after the merger between the EXPRESS and SOS workshops. We also used the collected data to compute a well-established measure of similarity between the two phases in the life of the SOS workshop, before and after the merger with EXPRESS. Beyond these data- and graph-mining analyses, we reflect on the major results developed in nearly four decades of research on SOS and identify, in our admittedly biased opinion, its strengths and gaps. 

The results of our quantitative and qualitative analyses all indicate a diminishing interest in the theory of SOS as a field of research. Even though `all good things must come to an end', we strive to finish this position paper on an upbeat note by addressing our second motivating question with some optimism. To this end, we use our personal reflections and an analysis of recent trends in two of the flagship conferences in the field of Programming Languages (namely POPL and PDLI) to draw some conclusions on possible future directions that may rejuvenate research on the (meta-)theory of SOS. We hope that our musings will entice members of the research community to breathe new life into a field of research that has been kind to three of the authors of this article. 

\paragraph{Whence this collaboration?} This article is the result of a collaboration between a researcher from the theory of algorithms and their applications, \href{https://www.pilucrescenzi.it/}{Pierluigi Crescenzi}, and three contributors to the theory of SOS. Pierluigi Crescenzi has recently offered data- and graph-mining analyses of conferences such as CONCUR, in cooperation with Luca Aceto in~\cite{AcetoCrescenzi2022}, SIROCCO~\cite{Crescenzi23} and ICALP---see the presentation available at \url{https://slides.com/piluc/icalp-50?token=fl3BBJ8j}. All authors thought that it was natural to combine quantitative data- and graph-mining analysis techniques with qualitative domain-specific knowledge to offer a fairly well-rounded perspective on the developments in the (meta-)theory of SOS and its relation to the SOS and EXPRESS/SOS workshops.  Both the Java code and the Julia software developed by Pierluigi Crescenzi, which was used for the quantitative analyses reported in this article and the aforementioned earlier ones, are publicly available at the following GitHub repository:  \url{https://github.com/piluc/ConferenceMining}. We encourage everyone interested in carrying out data- and graph-mining analyses of conferences to use it!

\section{Data Collection and Analysis}\label{Sect:data-analysis}

To set the stage for our reflections on the (meta-)theory of SOS, we have carried out some data analysis on the SOS and EXPRESS/SOS workshops. 

\subsection{Data Collection}
We extracted the following data from all the eleven past editions of the joint EXPRESS/SOS workshop:  
\begin{enumerate}
    \item the authors and titles of contributed talks; 
    \item invited speakers and the titles of their presentations or papers;   
    \item the number of submissions and accepted papers; and 
    \item at least two and at most three subject matter classifiers from the scope of EXPRESS/SOS.  
\end{enumerate}
Much of the gathered data was extracted from the tables of contents and proceedings of those editions of the workshop, which are all available in open access form as volumes of \href{https://cgi.cse.unsw.edu.au/~eptcs/}{Electronic Proceedings in Computer Science (EPTCS)}, and from the \href{https://dblp.org/db/conf/sos/index.html}{DBLP page devoted to the Workshop on Structural Operational Semantics}. In case of missing information regarding the number of submissions, we approached the workshops chairs and gathered that information through personal communication. For subject matter classification, since the general classifications, such as the one by the ACM, were too general for our purposes, we manually read the abstract (and in a few cases full papers) and identified domain-specific classifiers, using the scope definition of the EXPRESS/SOS workshop. 

The results of our data collection are publicly available \href{https://docs.google.com/spreadsheets/d/1OtDI1cbUV46cpT8W_l2BAUrmrXT5R8ZfDwFn2PHRuQ0/edit?usp=sharing}{online}.

The choice of focusing our analysis on the last eleven editions was motivated by the fact that, since 2012, the SOS workshop decided to join forces with the EXPRESS workshop and created a new joint venue. This gave us a consistent view of how the topics featured in the joint workshop have evolved over time and of how (structural) operational semantics has been represented in the joint workshop since 2012. However, using the data we collected, we also took the opportunity to compare the two phases of the SOS workshop, the first as an independent workshop in the period 2004--2011 and the second as EXPRESS/SOS from 2012 till 2022. 

\subsection{Automatic Analysis}

Based on the articles that were archived in the workshop proceedings, we found that 
\begin{itemize}
    \item 194 authors contributed articles to the workshop proceedings since 2004; 
    \item 90 colleagues published papers in the proceedings of the first eight editions of the SOS workshop; 
    \item 122 researchers contributed articles to the joint EXPRESS/SOS workshop in the period 2012--2022; 
    \item 18 authors published papers in the SOS workshop proceedings both before and after the merger with the EXPRESS workshop, which means that there were 104 contributors to EXPRESS/SOS who had never published in the SOS workshop in the period 2004--2011. 
\end{itemize} 
The above-mentioned data allow us to compute a measure of similarity between the two phases of the SOS workshop, before and after the merger with EXPRESS, using the S{\o}rensen-Dice index, which is a statistic used to measure the similarity of two samples. Given two sets $A$ and $B$, the \emph{Jaccard index} $J(A,B)$ is equal to $\frac{|A\cap B|}{|A\cup B|}$, and the \emph{S{\o}rensen-Dice index} is equal to $\frac{2J(A,B)}{1+J(A,B)}$, see~\cite{Dice,Soerensen}. 

The S{\o}rensen-Dice index for the lists of authors in the two phases of the SOS workshop is roughly $0.17$. This value indicates that the SOS workshop is not as similar to the joint EXPRESS/SOS workshop as one might have expected. By way of comparison, quoting from the data- and graph-mining analysis of CONCUR presented in~\cite{AcetoCrescenzi2022}, 
\begin{quote}
the conference that is most similar to CONCUR is LICS (with S{\o}rensen-Dice index approximately equal to $0.3$), followed by TACAS (approximately $0.25)$, CAV (approximately $0.24$), and CSL (approximately $0.21$).
\end{quote} 
Computing the S{\o}rensen-Dice index for SOS 2004--2022 and CONCUR, LICS, PLDI and POPL yields low values of similarity, namely $0.106396$ (CONCUR), $0.0622966$ (LICS), $0.00585138$ (PLDI) and $0.0303169$  (POPL). This is due to the fact that the sets of authors of those conferences is much larger than that of the SOS workshop, namely 1475 (CONCUR), 1953 (LICS), 3220 (PLDI) and 1979 (POPL). 

When quantifying the degree of similarity between a small workshop like SOS with larger conferences, it might be more appropriate to consider the {Szymkiewicz–Simpson coefficient} (also known as the overlap coefficient)~\cite{Simpson60,Szymkiewicz26,Szymkiewicz,SimSurvey}. 
Given two sets $A$ and $B$, the \emph{Szymkiewicz–Simpson coefficient} is equal to $\frac{|A\cap B|}{\min(|A|, |B|)}$.  
The values of that coefficient for the conferences we considered above are roughly $0.45$ (CONCUR), $0.34$ (LICS), $0.05$ (PLDI) and $0.17$ (POPL). Those values seem to support the view that SOS is rather similar to CONCUR and LICS, has some similarity with POPL, but is very dissimilar to PLDI. 

\subsection{Centrality Measures}

The \emph{static graph} (or collaboration graph) of SOS is an undirected graph whose nodes
are the authors who presented at least one paper at SOS, and whose edges link two authors who coauthored at least one paper (not necessarily presented
at SOS). In other words, this graph is the subgraph of the DBLP collaboration graph induced by the set
of SOS authors.

Centrality measures have been used as a key tool for understanding social networks, such as the static graph of SOS,  and are used to assess the
`importance' of a given node in a network---see, for instance,~\cite{Freeman78}. Therefore, to quantify the role played by authors who have contributed to the SOS workshop, we have
computed the following classic centrality measures on the largest connected component
of the static graph of SOS. 
\begin{itemize}
\item Degree: This is the number of neighbours of a node in the graph (that is, the number of coauthors).
\item Closeness: This is the average distance from one author to all other authors of its connected
component.
\item Betweenness: This is the fraction of shortest paths, passing through one author, between any
pair of other authors in its connected component.
\end{itemize}
The top ten SOS authors with respect to the above-mentioned three centrality measures are, in decreasing order:
\begin{itemize}
    \item Degree: Luca Aceto, Anna Ingólfsdóttir, Mohammad Reza Mousavi, Nobuko Yoshida, Rob van Glabbeek, Bas Luttik, Wan Fokkink, Michel Reniers, Catuscia Palamidessi, and Rocco De Nicola.
    \item Closeness: Luca Aceto, Rob van Glabbeek, Nobuko Yoshida, Matthew Hennessy, Ca\-tu\-scia Pa\-la\-mi\-des\-si, Anna Ingólfsdóttir, Rocco De Nicola, Daniele Gorla, Bas Luttik, and Uwe Nestmann.
    \item Betweenness: Luca Aceto, Matthew Hennessy, Nobuko Yoshida, Rob van Glabbeek, Rocco De Nicola, Catuscia Palamidessi, Daniele Gorla, Frank de Boer, Bartek Klin, and Uwe Nestmann.
\end{itemize}
In addition, we also calculated the \emph{temporal closeness}, which is an analogue of closeness that takes the number of years of a collaboration between two authors into account---see the paper~\cite{CrescenziMM20} for more information on this centrality measure. The top ten SOS authors according to temporal closeness are, in decreasing order: 
Luca Aceto, Anna Ingólfsdóttir, Wan Fokkink, Rocco De Nicola, Catuscia Palamidessi, Bas Luttik, Michel Reniers, Rob van Glabbeek, Jan Friso Groote, and  Mohammad Reza Mousavi.

Finally, to get a glimpse of the evolution of the aforementioned measures of similarity and centrality in the two phases of the SOS workshop, we computed them on the static graphs before and after the merger with EXPRESS. 

Before the merger with EXPRESS, the 2004--2011 editions of SOS had Szymkiewicz–Simpson index approximately of $0.42$ with CONCUR, $0.37$ with LICS, $0.067$ with PLDI and $0.2$ with POPL. After the merger with EXPRESS, those figures become  $0.512$ for CONCUR, $0.352$  for LICS, $0.032$ for PLDI and $0.152$ for POPL. So, from 2012 onwards, SOS has become more similar to CONCUR and even more dissimilar to PLDI and POPL than before. 

The top ten authors at the SOS workshop also change before and after the merger. When focusing on the period before the merger, the most central authors are as follows, in decreasing order:
\begin{itemize}
    \item {\sloppy Degree: Luca Aceto, Michel Reniers, Mohammad Reza Mousavi, Anna Ingólfsdóttir, Wan Fokkink, Rocco De Nicola, José Meseguer, Rob van Glabbeek, Catuscia Palamidessi, and David de Frutos-Escrig. \par}
    \item Closeness: Luca Aceto, Anna Ingólfsdóttir, Rocco De Nicola, Rob van Glabbeek, Matthew Hennessy, Georgiana Caltais, Mohammad Reza Mousavi, Eugen-Ioan Goriac, Michel Reniers, and Catuscia Palamidessi.
    \item Betweenness: 
Rocco De Nicola, Luca Aceto, Catuscia Palamidessi, José Meseguer, Frank de Boer, Filippo Bonchi, Matthew Hennessy, Michel  Reniers, Rob  van Glabbeek, and David de Frutos-Escrig.
\item Temporal closeness: 
Luca Aceto, Anna Ingólfsdóttir, Wan  Fokkink, Michel Reniers, Mohammad Reza Mousavi, José Meseguer, Jan Friso Groote, Rob van Glabbeek, Rocco De Nicola, and Ca\-tu\-scia Pa\-la\-mi\-des\-si.
\end{itemize}
After the merger with EXPRESS, our graph-mining analysis yields the following most central authors, in decreasing order:
\begin{itemize}
    \item Degree: Nobuko Yoshida, Luca Aceto, Bas Luttik, Rob van Glabbeek, Mohammad Reza Mousavi, Uwe Nestmann, Anna Ingólfsdóttir, Jorge Pérez, Jos Baeten, and Hans Hüttel.
    \item Closeness: Nobuko Yoshida, Luca Aceto, Rob van Glabbeek, Catuscia Palamidessi, Anna Ingólfsdóttir, Bas Luttik, Uwe Nestmann, Mohammad Reza Mousavi, Iain Phillips, and Mariangiola Dezani-Ciancaglini.
    \item Betweenness: Nobuko Yoshida, Rob van Glabbeek, Daniele Gorla, Luca Aceto, Bas Luttik, Bartek Klin, Uwe Nestmann, Catuscia Palamidessi, Hans Hüttel, and Rance Cleaveland. 
\item Temporal closeness: Luca Aceto, Anna Ingólfsdóttir, Bas Luttik, Tim Willemse, Ca\-tu\-scia Pa\-la\-mi\-des\-si, Mohammad Reza Mousavi, Jos Baeten, Jan Friso Groote, Jorge Pérez, and Rob van Glabbeek. 
\end{itemize}

\subsection{The Two Lives of the SOS Workshop}

\begin{figure}[ht]
    \centering
    \includegraphics[width=16cm]{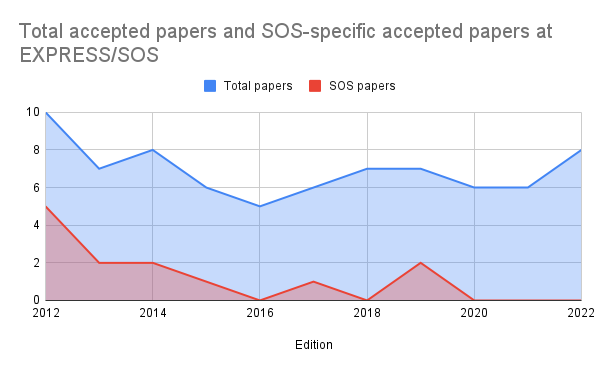}
    \caption{Total number of accepted paper (blue) and the number of accepted papers on SOS theory at the EXPRESS/SOS Workshop since 2012. }
    \label{fig:total-and-sos-papers}
\end{figure}

As we saw above, the first and the second life of the SOS workshop are not that similar after all, which seems to indicate that the eleven joint editions of the EXPRESS/SOS workshop were more about expressiveness than about structural operational semantics\footnote{Another possible explanation for the low degree of similarity between the pre- and post-merger incarnations of the SOS workshop is that the community welcomed many new authors from 2012 onwards. This would be a healthy and welcome development and is, in fact, supported by the data we collected. However, the analysis we present in what follows gives some indication that,  since 2014, the scientific programme of EXPRESS/SOS has featured only a few papers on structural operational semantics.}. To see whether this is really the case, we visually summarise the data we collected in Figure~\ref{fig:total-and-sos-papers} and provide its details below: 
\begin{itemize}
    \item The proceedings of EXPRESS/SOS 2012 included 10 papers, five of which dealt with topics related to operational semantics and its mathematical (meta-)theory---that's $50\%$  of the articles and the largest percentage of SOS contributions to EXPRESS/SOS in the period 2012--2022. 
    \item The proceedings of EXPRESS/SOS 2013 included seven papers, two of which dealt with topics related to operational semantics and its mathematical (meta-)theory---that's $28.5\%$ of the contributions .
    \item The proceedings of EXPRESS/SOS 2014 included eight papers, two of which ($25\%$) dealt with topics related to the theory of structural operational semantics.
    \item The proceedings of EXPRESS/SOS 2015 included six papers, one of which ($16.7\%$) dealt with topics related to the theory of structural operational semantics.
    \item The proceedings of EXPRESS/SOS 2016 included five papers, none of which dealt mainly with operational semantics.
    \item The proceedings of EXPRESS/SOS 2017 included six papers, one  of which ($16.7\%$) dealt mainly with operational semantics.
    \item The proceedings of EXPRESS/SOS 2018 included seven papers, none  of which dealt mainly with operational semantics.
    \item The proceedings of EXPRESS/SOS 2019 included seven papers, two  of which $28.5\%$ dealt mainly with operational semantics.
    \item The proceedings of EXPRESS/SOS 2020 included six papers, none of which dealt mainly with operational semantics.
    \item The proceedings of EXPRESS/SOS 2021 included six papers, none of which dealt mainly with operational semantics.
    \item The proceedings of EXPRESS/SOS 2022 included eight papers, none of which dealt mainly with operational semantics.
\end{itemize}

So, only 13 out of the 76 papers published in the proceedings of EXPRESS/SOS since 2012 dealt with topics in SOS theory ($17.1\%$ of published papers). In passing, we also note that 16 out of the 110 presentations  at the workshop in the period 2012--2022 were devoted to topics in SOS theory (that is, $14.5\%$ of the workshop presentations). Research in SOS was well represented at EXPRESS/SOS in the first three editions of the joint workshop.  However, five of the last seven  instalments  of the workshop did not include any presentations devoted to topics that were mainly related to structural operational semantics. In particular, EXPRESS/SOS 2020--2022 did not have any talks on the theory and applications of structural operational semantics.

\subsection{Reflections on the Analysis Results}
Reading through the EXPRESS/SOS contributions relevant to the theory of SOS reveals that the most recent results mostly focused on two aspects of SOS specifications: foundational aspects concerning the bialgebraic interpretation of SOS due to Turi and Plotkin~\cite{TuriP97}, as well as compositionality of quantitative notions of equivalence such as probabilistic bisimilarity. 
Below, we provide a more nuanced analysis of this trend.  

Another observation is that the diminishing strength in the provision of results on the theory of SOS can be largely attributed to a lack of projects (particularly, PhD~studentships) in this area. Almost all of the results on the meta-theory of SOS contributed to the EXPRESS/SOS series had a co-author with a PhD~project on this topic. A reduction in the number of doctoral students does not bode well for the healthy development of any research field.  

\section{Personal Reflections}\label{Sect:reflections}

Since the appearance of Plotkin's seminal Aarhus technical report~\cite{Plotkin81}, reprinted in slightly revised form as a journal paper in~\cite{Plotkin04a} with some historical remarks by Plotkin himself in~\cite{Plotkin04}, structural operational semantics has arguably become the most widely used approach to defining the semantics of programming and executable specification languages. To our mind, it is as impactful and popular today as it has been for over forty years. Indeed, one would be hard pressed to find papers on the theory of programming and specification languages that do not use structural operational semantics in some way. Moreover, the semantics of full-blown programming or domain-specific languages is still given in that style, reflecting its flexibility and applicability---see, for instance, the paper~\cite{GrishchenkoMS18} for a small-step semantics of full Ethereum-virtual-machine bytecode that is formalised in the $F*$ proof assistant~\cite{mumon} and then validated against the official Ethereum test suite.

As Plotkin highlights in his aforementioned piece on the origins of structural operational semantics, the essence of that approach to semantics is that it is \emph{rule based} and that the rules should be \emph{syntax directed} in order to support compositional language specifications and reasoning, as in the denotational approach to semantics. Conceptually, this rule-based view of operational semantics naturally led to the development of a theory of SOS language specifications that focused on the rules used in semantic definitions. The gist of that line of research, which can be traced back to de Simone's work~\cite{deSimone85}, was to study \emph{rule formats} for operational specifications guaranteeing that every program in the specified language afford some semantic property of interest. So, rule formats offered a way to reduce the checking of semantic properties of programs in a language to syntactic checks on the rules used to define the operational semantics of the language. The literature on what came to be called the `meta-theory of structural operational semantics' is by now very large and we cannot do it justice in this paper. We refer the interested reader to the survey articles~\cite{AcetoFV01,MousaviRG07} and to the references therein as well as the proceedings of SOS, EXPRESS/SOS, and of conferences such as CONCUR, LICS and POPL, for much more information and recent references. Naturally, since its first edition in 2004, the SOS workshop has served as a venue for the publication of several articles on SOS meta-theory.

Three of the authors of this piece have been amongst the contributors to the development of the fascinating research on rule formats for operational specifications and thoroughly enjoyed doing so. However, we feel that the time has come for a critical appraisal of the strengths, weaknesses and possible future of that line of research and to speculate about whether the data we discussed in Section~\ref{Sect:data-analysis} reflects the musings we present in the rest of this note.

\subsection{Strengths}

In our, admittedly biased, opinion, research on rule formats for structural operational semantics has led to a wealth of interesting and elegant theoretical results, ranging from those on the meaning of rule-based specifications using rules with negative premises (see, for instance, the articles~\cite{BolG96,Glabbeek04,ChurchillMM13}) to congruence formats for several behavioural equivalences obtained uniformly from their modal characterisations via modal decomposition (see, for example,~\cite{BloomFG04,FokkinkGW12,FokkinkG17,FokkinkGL19} and the references therein). Early studies of congruence rule formats, such as those reported in the seminal~\cite{BloomIM95,GrooteV92}, were accompanied by characterisations of the largest congruences included in trace semantics induced by the collection of operators that can be specified in the rule formats studied in those references. After all these years, we still find it amazing that such results could be proved at all!

Below we provide a non-exhaustive list of the available meta-theorems with sufficient strength (more than a single paper, with more than one application to a language) and we refer to the past review papers/chapters ~\cite{AcetoFV01,MousaviRG07}  for a more exhaustive list to the date of their publication:

\begin{itemize}
    \item Congruence: proving congruence (compositionality) for various notions of strong \cite{Middelburg01,Verhoef95}, weak \cite{FokkinkG17}, higher-order \cite{MousaviGR05}, data-rich \cite{MousaviRG05}, timed \cite{Kick02}, and quantitative behavioural equivalences \cite{DArgenioGL16,CastiglioniGT18,CastiglioniT20}; supporting various syntactic language features such as formal variables and binders \cite{Middelburg01,CiminiMRG12}, as well as semantic features such as negative premises and predicates, terms as labels, and ordering on rules.

    \item (De-)Compositional reasoning methods: decomposing logical formulae (in the multi-modal $\mu$-calculus, also known as Hennessy-Milner logic with recursion,~\cite{Kozen83,Larsen90}) according to the semantics of various operators for various notions of bisimilarity \cite{FokkinkGL19,FokkinkG17,FokkinkGW12} and their quantitative extensions \cite{CastiglioniGT18,CastiglioniT20}; interestingly, this can lead not only to a reasoning method for checking modal formulae, but can also serve as a recipe for `generating' congruence formats for different notions of equivalence, once their modal characterisation is defined.

    \item Axiomatisation and  algebraic properties: to generate sound and ground-complete axiomatisations for strong bisimilarity \cite{DBLP:journals/iandc/AcetoBV94}, as well as weak behavioural equivalences \cite{Glabbeek11}, and equivalences with data \cite{GeblerGM13}. An orthogonal line of enquiry considered identifying sufficient conditions guaranteeing  various algebraic properties of language operators such as commutativity \cite{MousaviRG05a}, associativity \cite{CranenMR08}, zero and unit elements \cite{AcetoCIMR11}, and idempotence \cite{AcetoBIMR12}; we refer to an accessible overview paper \cite{AcetoIMR09} summarising such results to its date of publication. 
\end{itemize}
There have been a number of implementations of such results in tools \cite{AcetoGI13,MousaviR06,VerdejoM06}, mostly based on rewriting logic \cite{Maude2007}.

Several of the theorems from the theory of structural operational semantics have found application in the study of process calculi, reducing the need to prove congruence and axiomatisation results, amongst others, from scratch for each calculus and have been extended to settings including, for instance, probabilistic and stochastic features (see, for example,~\cite{CastiglioniT20,DArgenioGL16}), as well as to higher-order calculi, as in the recent~\cite{GoncharovMSTU23}. The article~\cite{GoncharovMSTU23} belongs to a fruitful and still active line of research, stemming from the seminal work by Turi and Plotkin~\cite{TuriP97}, providing bialgebraic foundations to the theory of structural operational semantics.

The contributions to the work on rule formats and on the meta-theory of structural operational semantics have striven to achieve a reasonably good trade-off between the generality of the technical results and the ease with which they can be applied to specific languages. Ideally, one would always like to have simple syntactic restrictions on rule formats that guarantee desired semantic properties in a wide variety of applications. Indeed, following a Pareto Principle, very often simple rule formats cover many of the languages of interest and one quickly hits a threshold where complex and hard-to-check definitions are needed to extend the applicability of obtained results. In many cases, the `curse of generality' led to definitions of rule formats whose constraints are arguably not purely syntactic any more and may even be undecidable. As an example, Klin and Nachyla ~\cite{KlinN17} have shown  that it is undecidable whether an operational specification that uses rules with negative premises has a least supported model and whether it has a unique supported model or a stable model. It is also undecidable whether such a specification is complete. As mentioned by Klin and Nachyla in the aforementioned reference, these negative results entail that formats such as the complete ntyft/ntyxt~\cite{FokkinkG96} `are not \emph{bona fide} syntactic
formats, as there is no algorithmic way to tell whether a given specification fits such a format.' So, the pursuit of generality is, to our mind, a double-edged sword and can be seen as both a strength and a weakness of several result on rule formats and the meta-theory of structural operational semantics.

In the context of EXPRESS/SOS, we observed that this tradition of strong theoretical results is dying down: from 2012 to 2017, we counted nine contribution to the foundation of SOS specifications \cite{AcetoGI13,Bonsangue2012,DArgenioLG15,GeblerGM13,GeblerT13,GeblerT14,KlinN17,Lee2012,Rot2017}, including on the bialgebraic framework ~\cite{Bonsangue2012,KlinN17,Rot2017}, as well as congruence for quantitative notions of equivalence \cite{DArgenioLG15,GeblerT13,GeblerT14,Lee2012} and axiomatisation results \cite{GeblerGM13}; however, this number dropped to only one contribution from 2018 to 2022 on the meaning of SOS specification and compositionality of equivalences on open terms  \cite{vanGlabbeek2019}. 

In summary, we believe that the study of rule formats and of the meta-theory of structural operational semantics has yielded many elegant results that have been of some use for the working concurrency theorist. However, first, the number of such contributions has significantly dropped in the past few years and, second, one has to wonder whether that line of work has had impact on the field of programming language semantics. We will offer some musings on that question in the coming section.

\subsection{Gaps}

To our mind, apart from its intrinsic scientific interest, the theory of  structural operational semantics based on rule formats has served the concurrency-theory community well by providing elegant, and often general and deep, results that have both explained the underlying reasons why specific languages enjoyed several semantic properties and served as tools to prove new theorems as instances of a general framework. The use of `syntactic' rule formats to establish properties of interest about formal systems has also been used in logic. By way of example, Ciabattoni and Leitsch have given algorithmic conditions guaranteeing that some logics enjoy cut elimination ~\cite{CiabattoniL08}. However, despite its undoubted successes, to our mind, the theory of rule formats has not yet had the impact one might have expected on the community working on the theory of  programming languages. Perusing the proceedings of the premier conferences in that field indicates that much of the research on programming-language semantics and its applications is done in the context of proof assistants such as Coq~\cite{BertotC04,CoquandH88}\footnote{Coq is available at \url{https://coq.inria.fr/}.} and on frameworks built on top of those---see, for instance, the highly influential Iris framework for higher-order concurrent separation logic ~\cite{JungKJBBD18}. 

We speculate that this relative lack of impact might be due to the fact that the theory of structural operational semantics based on rule formats has been mostly developed within the process algebra community. This has naturally led to the development of results and frameworks having process calculi as main application area. As a consequence, despite some foundational research~\cite{AcetoFGIO19,FioreS09,MousaviRG05}, the development of a widely-applicable theory of rule formats for languages having first-class notions of data and memory, as well as binding constructs is still somewhat in its infancy. This limits the applicability of the results obtained by the concurrency theory community to mainstream programming languages. Moreover, the software tools embodying the theory of structural operational semantics developed so far have mostly taken the form of prototypes and are arguably not as mature and usable as those produced by groups working on the theory of programming languages \cite{SewellNOPRSS10}. The initial work carried out within the PLanCompS ~\cite{BinsbergenSM16} aimed to address this gap based on the Modular SOS framework that has been pioneered by Mosses~\cite{Mosses04}; this line of work has been influential and has led to other frameworks such as the iCoLa framework for incremental language development \cite{FrolichB22}.

\subsection{Trends and Opportunities}

To relate the past strengths to future trends, particularly regarding emerging application areas of operational semantics, we analysed the table of contents of five past editions of flagship conferences in programming languages: POPL (from 2021 to 2023, inclusive) and PLDI (from 2021 to 2022, inclusive). The aim of the analysis was to find areas where the available strength in the theory of SOS can be exploited. We aimed to be as inclusive as possible and tried to mention any such areas, even if the exploitation of available strength would require a major rework or transformation of ideas and results. Below we provide a raw list of keywords that we encountered in our analysis:  

\begin{itemize} 
\item POPL 2023: Semantics of Probabilistic and Quantum programs, Coq Proof Libraries, Nominal Sets,  Co-Algebra and Bisimulation, Multi-Language Semantics, Session types.

\item  POPL 2022: Session types, Semantics of Probabilistic and Quantum programs, Semantic Substitution and congruence. 

\item POPL 2021:  Semantics of Probabilistic Programs, Nominal Semantics, Hyper-properties and non-interference, functorial semantics

\item PLDI 2022: Information flow analysis, equational and algebraic reasoning (also applied to quantum programs), sound sequentialisation, Kleene algebra, language interoperability, verified compilation (also applied to quantum programs).

\item PLDI 2021: Language translation conformance and compiler verification, 
session types, regular expressions, semantics of probabilistic and quantum programs.
\end{itemize}

In all the POPL and PLDI editions we reviewed, abstract interpretation (also for quantum programs), analysing weak memory models, and reasoning using separation logics are featured prominently.  

It appears from our analysis that the following activities may have substantial potential impact:

\begin{itemize} 
\item semantic meta-theorems about quantitative transition systems (particularly probabilistic and quantum transition systems \cite{BornatBKPN20,FengDY14});

\item providing mechanised semantic frameworks, particularly in proof assistants such as Coq; 
\item defining general semantic frameworks and theorems for different memory models and models of parallelism; 
\item defining general compositional frameworks for reasoning with separation logics and logics of incorrectness; 
\item devising algorithms for test-case generation, for instance, for compiler testing, based on a semantic framework.  
\end{itemize}
We hope to see work on some of those topics in the near future, which might lead to a new lease of life for the (meta-)theory of SOS and its applications.

\paragraph{Acknowledgements} We thank Valentina Castiglioni and Peter Mosses for their comments on a draft of this piece. Luca Aceto and Anna Ing\'olfsd\'ottir were partly supported by the projects  `Open Problems in the Equational Logic of Processes (OPEL)' (grant no.~196050) and `Mode(l)s of Verification and Monitorability (MoVeMent)' (grant no.~217987) of the Icelandic Research Fund. Mohammad Reza
Mousavi have been partially supported by the UKRI Trustworthy Autonomous Systems Node in Verifiability, Grant
Award Reference EP/V026801/2 and the EPSRC grant on  Verified Simulation for Large Quantum Systems (VSL-Q), Grant Award Reference EP/Y005244/1.

\bibliographystyle{eptcs}
\bibliography{generic}
\end{document}